\RequirePackage{fix-cm}
\documentclass[smallextended]{svjour3}       % onecolumn (second format)
\smartqed  % flush right qed marks, e.g. at end of proof
\usepackage{graphicx}
\usepackage{bm}        % for math
\usepackage{amssymb}   % for math
\usepackage{amsmath}
\usepackage{color}
\usepackage{subfig}
\begin{document}

\title{Consequences of the Inherent Density Dependence in Dirac Materials}

\author{Matthew P. Gochan         \and
        Kevin S. Bedell %etc.
}


\institute{M.P. Gochan \at
              Department of Physics \\
              Boston College\\
              140 Commonwealth Avenue\\
              Chestnut Hill, MA, 02467\\
              \email{gochan@bc.edu}           
}

\date{Received: date / Accepted: date}

\maketitle

\begin{abstract}
Dirac materials are systems in which the dispersion is linear in the vicinity of the Dirac points. As a consequence of this linear dispersion, the Fermi velocity is independent of density and these systems exhibit unusual behavior and possess unique physical properties that are of considerable interest. In this work we study the ground state behavior of 1D Dirac materials in two ways. First, using the Virial Theorem, we find agreement with a previous result in regards to the total average ground state energy. Namely, that the total average ground state energy, regardless of dimensionality, is found to be $E=\mathcal{B}/r_s$ where $\mathcal{B}$ is a constant independent of $r_s$. As a consequence, thermodynamic results as well as the characteristic exponents of 1D Fermi systems are density independent. Second, using conventional techniques, i.e. Tomanaga Luttinger theory (TLL), we find several unique properties that are a direct consequence of the dispersion. Specifically, the collective modes of the system exhibit electron density independence predicted from the Virial Theorem. Finally, experimental techniques are discussed in which these  predictions of density independent exponents and their behavior can be tested. 

\end{abstract}

\section{Introduction}
\label{intro}
Dirac materials encompass a large family of materials that are at the forefront of theoretical and experimental research. Graphene, topological insulators, and \textit{d}-wave superconductors are a few examples in this family that have been thoroughly studied and exhibit new results and rich physics. These systems, in spite of being different, share a fundamental similarity that links them together: their low energy fermionic excitations behave as massless Dirac particles \cite{Dirac}. A standard tight-binding calculation immediately leads to an exact linear dispersion in the vicinity of the Dirac points and is vastly different from the parabolic dispersion found in ordinary systems. Thus, any properties that are a direct consequence of this linear spectrum are universal among these diverse materials. For example, the fermionic specific heat is controlled by the linear behavior and its power law dependence is shared by graphene, \textit{d}-wave superconductors, and superfluid $^3$He. In addition to these low energy properties, some Dirac materials are also boundary states for exotic quantum phases and have drawn great research interest. Specifically, the so-called Dirac semi-metal can be driven into various quantum states such as Weyl semimetals, axion insulators, and topological superconductors \cite{Liu}. Although Dirac materials possess different properties. It is the universal low energy behavior, the possibility of realizing exotic quantum states of matter, and the rich fundamental physics within Dirac materials that has fueled the research into understanding these systems.

The system used to model a 1D Dirac material is a single walled carbon nanotube (SWNT) in its metallic state at $T=0$ where we stay away from gapped states and avoid possible Wigner crystalliztion  by avoiding other chiralities \cite{Deshpande,DeshBock}. The Hamiltonian is (we set $\hbar=1$ for the remainder of the paper)\cite{Solyom} 
\begin{equation}
H=H_0+H_{int}
\end{equation}
where $H_0$ is the Hamiltonian for a free Dirac gas:
\begin{equation}
H_0=v_g\sum_{r,k,\sigma}\left(rk-k_F\right)c_{k\sigma}^{r\dagger}c_{k\sigma}^r
\end{equation} 
and $H_{int}$ is the Tomanaga-Luttinger interaction \cite{Krastan}:
\begin{equation}
\begin{aligned}
H_{int}=\frac{1}{L}\sum_{k_1,k_2,p,\alpha,\beta}
\left[\Gamma_{\alpha,\beta}^2c_{k_1,\alpha}^{+\dagger}c_{k_2,\beta}^{-\dagger}c_{k_2+p,\beta}^-c_{k_1-p,\alpha}^{+}\right.+&\frac{1}{2}\Gamma_{\alpha,\beta}^4\left(c_{k_1,\alpha}^{+\dagger}c_{k_2,\beta}^{+\dagger}c_{k_2+p,\beta}^+c_{k_1-p,\alpha}^+\right.\\
&\left.\left.+c_{k_1,\alpha}^{-\dagger}c_{k_2,\beta}^{-\dagger}c_{k_2+p,\beta}^-c_{k_1-p,\alpha}^-\right)\right]
\end{aligned}
\end{equation} 
where the Fermi velocity $v_g\sim10^6$ m/s is a constant, and $c^{\dagger},c$ are the creation and annihilation operators for fermions. \footnote{$\Gamma_{\alpha,\beta}^i=g_i^s\delta_{\alpha,\beta}+g_i^a\delta_{\alpha,-\beta}$ $(i=2,4)$ describes the interaction between electrons; $i=2$ refers to forward scattering between electrons on separate branches while $i=4$ refers to forward scattering between electrons on the same branch. The superscripts \textit{s} and \textit{a} denote symmetric and anti-symmetric spin respectively.} The coupling constants, $g_i^s$ and $g_i^a$, are a measure of the interaction strength between electrons for given scattering processes and are in general momentum dependent; we neglect this dependence by restricting the momentum to within certain values $\Lambda$ \cite{Solyom}. \footnote{The interaction given in (3) only contains terms involving forward scattering; Umklapp and back scattering processes have been ignored. This exclusion is justified through the following: as long as we're considering low-energy properties, interactions that don't commute with charge and spin separation are irrelevant \cite{Voit}.} It is important to note that the interaction in (3) is not complete but is the simplest model that shows deviation from typical Fermi liquid behavior. 

We immediately notice an important distinction in (2) unique to 1D Dirac materials that isn't seen in conventional 1D metals; although TLL theory produces exact results for the Green's function in the vicinity of the fermi points ($\pm k_F$), the result is hinged upon the linearity of the dispersion relation \cite{Tomonaga,Luttinger}. In ordinary 1D metals, this linear behavior is obtained by linearizing the quadractic dispersion around $\pm k_F$ and is therefore only exact in a region within a momentum cutoff (e.g. $k_F-\Lambda<k<k_F+\Lambda$). In Dirac materials, the linear behavior exists naturally without linearization leading to validity for all \textit{k}.
\begin{figure}[h]
  \centering
  \includegraphics[scale=0.5]{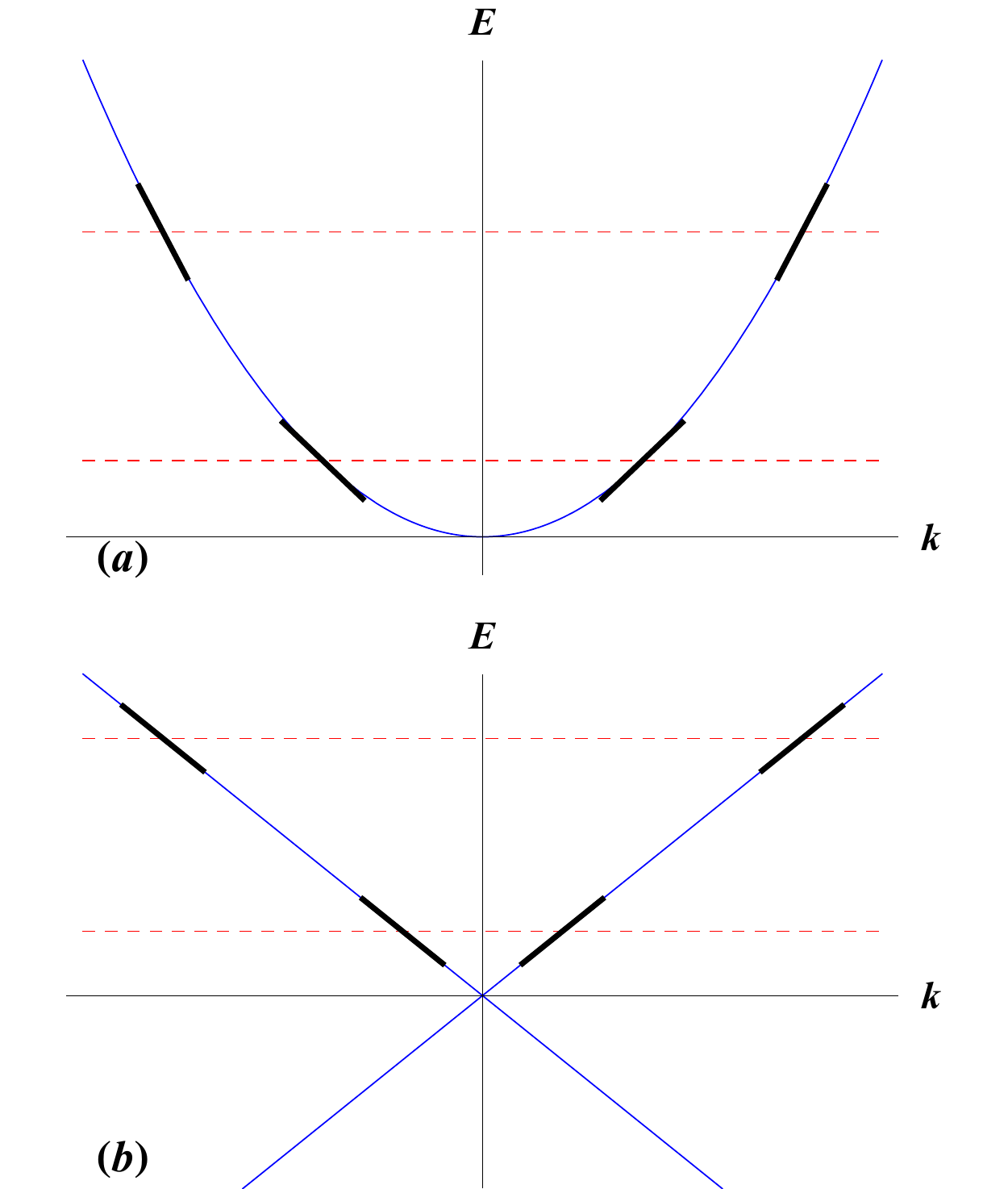}
  \caption{(color online) \small{(a) The dispersion for a normal 1D metal. The red dashed lines denote different values for the Fermi energy. Notice how the linearization at different $k_F$ points (i.e. different density) produces different slopes (black lines) and therefore different values for $v_F$. (b) The dispersion for a 1D Dirac metal. No matter where the Fermi energy is, the slope, and therefore $v_F$, is the same.} }
\end{figure}

Recent angle resolved photoemission spectroscopy (ARPES) measurements performed on SWNTs have exposed behavior consistent with TLL theory \cite{Ishii}. Such behavior, which is vastly different from traditional Fermi-liquid Theory behavior, was seen in the spectral function and intensities which both exhibit the desired power-law behavior in good agreement with theory. These results motivated our investigation of SWNTs and lead to the main question: will the unusual behavior of electrons in SWNTs lead to deviations from TLL behavior? As we will see, a number of non-trivial results emerge in SWNTs that aren't seen in ordinary 1D metals. Such differences allude to a new class of low dimensional interacting systems which we call Dirac Liquids.

This paper is outlined as follows: First we derive the total ground state energy of the system using the Virial Theorem and discuss the implications of the result as well as the apparent connection to Coulomb systems. Second we use TLL theory to derive numerous quantities, for example the density of states, and explicitly show this density independence. Next, we discuss potential experiments that can test our predictions. Finally, concluding remarks.

\section{Average Ground State Energy}
\label{Virial}
Recent work on Dirac materials in the presence of a Coulomb potential using the Virial Theorem provides an expression for the average ground state energy of the system \cite{Dusty}:
\begin{equation}
E\left(r_s\right)=\frac{\mathcal{B}}{r_s}
\end{equation}
where $\mathcal{B}$ is a constant independent of density. In this paper, $r_s$ is a dimensionless constant that's a measure of density. Specifically, $r_s$ allows us to express the electron spacing in units of a characteristic length of the system; in this case, we have $r=ar_s$. Although the work in \cite{Dusty} was done for 2D, the form for (4) remains the same regardless of the dimension of the system; the only thing that changes with a change in dimension is the constant $\mathcal{B}$. Therefore, we expect the average ground state energy of a 1D Dirac material to have the same form as (4).

We start with an expression relating the compressibility of the system to the average ground state energy:
\begin{equation}
\kappa^{-1}=n^2\frac{\partial^2\mathcal{E}}{\partial n^2}
\end{equation}
where $\mathcal{E}=E/l$. The benefit in using (5) is the following: in 1D, the expression $n^2\kappa$ is independent of density. This allows for a straightforward integration to obtain $\mathcal{E}$: 
\begin{equation}
\mathcal{E}=\frac{1}{2}\kappa^{-1}
\end{equation}
where the compressibility for a 1D interacting system with Dirac spectrum is
\begin{equation}
\kappa=\frac{4v_g}{\pi n^2\left(v_1\right)^2}
\end{equation}
where $v_1$ is the speed of sound. The total average ground state energy is then:
\begin{equation}
E\left(r_s\right)=\frac{N\pi v_g}{8a}\left(1+\frac{2g_4}{\pi v_g}\right)\frac{1}{r_s}
\end{equation}
where $a=1.44$ \AA \cite{Pines,Saito}. The term in front of $1/r_s$ is  
\begin{equation}
\frac{\mathcal{B}}{N}=\frac{\pi\hbar v_g}{8a}\left(1+\frac{2g_4}{\pi v_g}\right)
\end{equation}
(where $\hbar$ has been restored) and, as expected, is independent of $r_s$. For a typical SWNT, $\frac{\mathcal{B}}{N}\approx 2.8\pm.2$ eV.

Equation (8) is the main result in this paper and is nontrivial with a few remarkable consequences: first, there seems to be a connection between a Tomanaga-Luttinger system and a Coulomb system. Previous work in \cite{Egger,KBF} have explored SWNTs in the presence of a long-range Coulomb potential. These systems are vastly different, yet their average ground state energies have the same form and dependence on $r_s$, with the exception that the constant $\mathcal{B}$ varies slightly. Such a similarity alludes to the specific form of the potential not being as important as the linear dispersion in controlling the underlying physics of the system. Second, a closed form for the ground state energy (8) provides a bound allowing calculation of the thermodynamics and dynamics of a 1D Dirac material. This closed form is a rare property that appears in all Dirac materials regardless of dimension \cite{Dusty}. 

\section{Tomanaga-Luttinger Result}

Typically, interacting Fermi systems are examined through the lens of Fermi Liquid Theory. Unfortunately, for a 1D system, Tomanaga-Luttinger Liquid theory is necessary to describe the low-energy asymptotic behavior since the excitations are collective excitations (boson modes) rather than single particle excitations \cite{Solyom,Voit}. Within this formalism, we first derive the Green's function in a similar way that Dzyaloshinskii and Larkin have \cite{Dzyalo}. Ignoring interactions, the Green's functions for each branch is
$$
G_r^{(0)}\left(k,\omega\right)=\frac{1}{\omega-v_g\left(rk-k_F\right)+i\delta\text{sgn}\left(rk-k_F\right)}\hspace{.25in}r=+,-
$$
In typical problems, the vertex function $\Gamma$ is determined via an infinite series of diagrams and requires approximations. In the model under consideration, there exists a simple and exact relation between $\Gamma$ and $G_r(k,\omega)$ and thus for each branch is 
\begin{equation}
\Gamma_r=\frac{G_r^{-1}\left(p,\varepsilon\right)-G_r^{-1}\left(p-k,\varepsilon-\omega\right)}{\omega-rv_gk}
\end{equation}
and is a direct consequence of the spectrum given by eqn (2) and conservation of particle number on each branch. Eqn (10) allows the Dyson equation \footnote{Diagrammatic representations for Dyson's equation as well as the vertex function can be found in \cite{Dzyalo} and \cite{Solyom}} to be expressed as
\begin{equation}
\left[\varepsilon-v_g\left(p-k_F\right)\right]G_+\left(p,\varepsilon\right)=1+\frac{i}{4\pi^2}\int\frac{D_{4\parallel}}{\omega-v_gk}G_+\left(p-k,\varepsilon-\omega\right)
\end{equation}
where $D_{4\parallel}$ \footnote{As a consequence of the linear dispersion, diagrams which contain loops with more than two interaction vertices cancel out and all that's left is self energy diagrams containing bubbles. These are summed into the effective interactions. Expressions for the effective interactions for a normal metal can be found in \cite{Solyom}} is the effective coupling constant \cite{Solyom}. Now we Fourier transform eqn (11) to get  
\begin{equation}
\begin{aligned}
G_r(x,t)=\frac{1}{2\pi}\frac{rx-v_gt+i/\Lambda}{rx-v_gt+i\delta}&\prod_{j=\sigma,\rho}\frac{1}{\left(rx-u_jt+i/\Lambda\right)^{1/2}}\\
&\frac{1}{\left[\Lambda^2\left(x^2-u_j^2t^2+\frac{i}{\Lambda}\left(2u_jt-\frac{i}{\Lambda}\right)\right)\right]^{\alpha_j}}
\end{aligned}
\end{equation} 
where $\delta$ is the bandwidth cutoff and $\Lambda$ is the momentum transfer cutoff. $u_\sigma$ and $u_\rho$ are the velocities of the spin density and charge density bosonic mode respectively; they depend on $v_g$, $g_2$ and $g_4$. The exponents, $\alpha_\sigma$ and $\alpha_\rho$, are of special interest and have the following form:
\begin{subequations}
	\begin{align}
	\alpha_\sigma=\frac{1}{4u_\sigma}\left[v_g+\frac{1}{2\pi}\left(g_4^s-g_4^a\right)-u_\sigma\right] & \\
	\alpha_\rho=\frac{1}{4u_\rho}\left[v_g+\frac{1}{2\pi}\left(g_4^s+g_4^a\right)-u_\rho\right]
	\end{align}
\end{subequations}

$\alpha_\sigma$ and $\alpha_\rho$ are a measure of interaction strength in the system and are the signature of 1D interacting Fermi behavior. The exponents for a normal 1D metal can be found in \cite{Solyom,Voit} and although seem similar, are different in a subtle, yet crucial, way. The dependence of the Fermi velocity on electron density is linear in a 1D normal metal, i.e. $v_F\propto n$ as seen in Fig. (1a). This leads to different values for equilibrium and dynamic quantities dependent on which \textit{k}-point has been linearized around. This is vastly different in a 1D Dirac metal. The Fermi velocity, $v_g$, is constant in electron density as seen in Fig. (1b). This leads to our result being exact, and equilibrium and dynamic properties are independent of $k_F$. Additionally, this difference has consequences in understanding the interactions in the 1D Dirac materials. Since (13a) and (13b) are density independent, the interaction strength in these systems is constant regardless of particle number. 

From (12) we derive the momentum distribution function as done in \cite{Voit2} using the following equation 
\begin{equation}
n(k)=-i\sum_r\int_{-\infty}^{\infty}G_r(x,0^-)e^{-ikx}dx
\end{equation}
At $t=0$, the Green's function behaves as
$$
G_r(x)\sim x^{-1-\alpha}\hspace{.5in}\alpha=2\left(\alpha_\sigma+\alpha_\rho\right)
$$
we therefore expect the momentum distribution to have the same form. Using the Green's function given by eqn (10), we arrive at the following expressions:
$$
\begin{aligned}
n(k)=&-\frac{1}{2\pi}\int_{-\infty}^{\infty}\frac{dx}{\alpha-irx}\left(1+\left(\frac{x}{\Lambda}\right)^2\right)^{\alpha/2}e^{-i\left(k-rk_F\right)x}\\
=&\frac{1}{\pi}\text{sgn}\left(k_F-rk\right)\int_0^\infty\frac{dx}{x}\left(1+\left(\frac{x}{\Lambda}\right)^2\right)^{\alpha/2}\sin\left(\left|rk-k_f\right|x\right)
\end{aligned}
$$
where the limit $a\rightarrow 0$ has been taken in the second equality. Evaluation of the integral is done with \cite{GR} and the general solution contains a combination of special functions. Upon taking the appropriate limits\footnote{Since we are interested in the asymptotic low energy behavior, we have the following limits for the validity of $n(k)$:
\begin{enumerate}
	\item $\Lambda^2\rightarrow0$
	\item $\left|rk-k_F\right|^2\rightarrow0$
\end{enumerate} }, the final result for the momentum distribution is 
\begin{equation}
n(k)\simeq\frac{1}{2}-C_1\left|k-rk_F\right|^\alpha\text{sgn}\left(rk-k_F\right)-C_2\left(k-r k_F\right)
\end{equation}
where $C_1$ and $C_2$ are constants that depend on $\alpha$ and $\Lambda$\footnote{The coefficients are
\begin{enumerate}
	\item $C_1=\frac{1}{2\sqrt{\pi}}\frac{\Gamma\left(\frac{1}{2}-\frac{\alpha}{2}\right)}{\Gamma\left(1+\frac{\alpha}{2}\right)}\left(\frac{\Lambda}{2}\right)^{\alpha}$
	\item $C_2=\frac{\beta\left(\frac{1}{2},\frac{1}{2}\left(\alpha-1\right)\right)}{2\pi}\Lambda$
\end{enumerate}. In following with the theme, these coefficients are independent of electron density}. Eqn (13) exhibits all the usual behavior expected in a 1D system; specifically, the continuity of $n(k)$ as well as the singularity in the first derivative indicating failure of the quasi-particle picture.

With the momentum distribution (13), we derive the density of states: 
$$
g(E)=\frac{1}{L}\frac{dN}{dE}
$$
where \textit{L} is the length of the SWNT and \textit{N} is the number of electrons given by:
$$
N=\frac{L}{2\pi}\int n(k)dk
$$
We limit ourselves to a small area, $\delta k$, around $\pm k_F$ where eqn. (13) is valid. In this area, small changes in \textit{N} with respect to \textit{k} can be expressed as: 
$$
\frac{\delta N}{\delta k}=\frac{L}{2\pi}\left(1-\frac{C_1}{\alpha+1}\left(\delta k\right)^\alpha+\frac{C_2}{2}\delta k\right)
$$
In the limit as $\delta k\rightarrow 0$, the $\delta k$ term is negligible since $0<\alpha<1$ \cite{Ishii,Bockrath}. Finally, we obtain the density of states for a 1D Dirac material near $\pm k_F$:
\begin{equation}
g(\omega)=\frac{2}{\pi v_g}\left(1-\frac{C_1}{v_g^\alpha\left(\alpha+1\right)}\left|\omega\right|^\alpha\right)
\end{equation}
where $C_1$ is the same as it is in eqn (13) and $\omega$ is measured with respect to $\pm k_F$. Our expression agrees with the ARPES results in \cite{Ishii} that fit their data with $g(\omega)\sim|\omega|^\alpha$.  Additionally, (14) reduces to the expression for the density of states for a non-interacting 1D Dirac system ($\alpha=0$).

\begin{figure}[h]
\centering
\includegraphics[scale=.5]{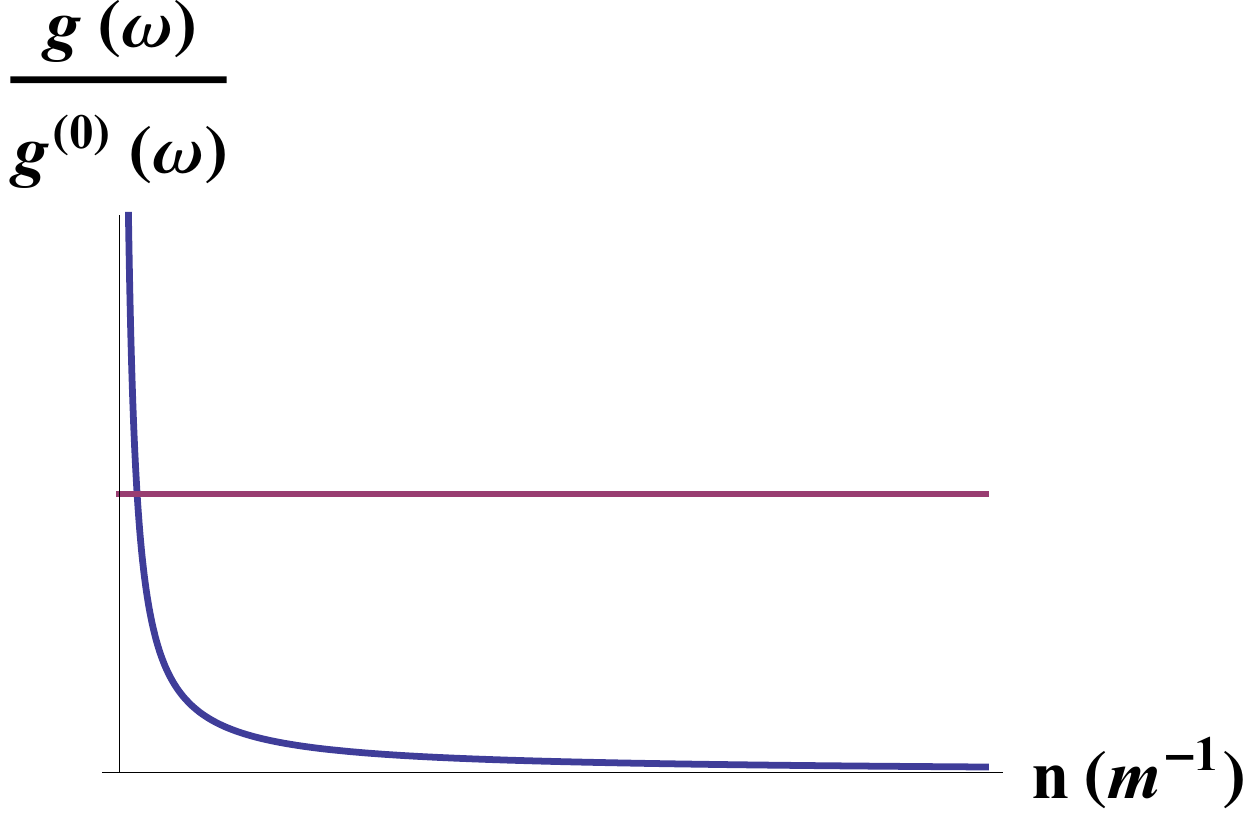}
\caption{(color online)\small{ Plot of density of states normalized to the non-interacting value as a function of electron density. The blue curve represents $g/g^{(0)}$ for a normal metal while the purple line represents $g/g^{(0)}$ for a Dirac system. We see a clear distinction in the behavior of the density of states between the systems indicating that the dimensionality of the system, although important, isn't nearly as important as the behavior of the dispersion.}}
\end{figure}
The unique features of eqn (14) are displayed in figures 2-4. In fig 2 we see clear distinction between the 1D normal metal and the 1D Dirac metal in terms of the dependence on electron density. The result implies a saturation of the density of states.
\begin{figure}[h]
\centering
\includegraphics[scale=.5]{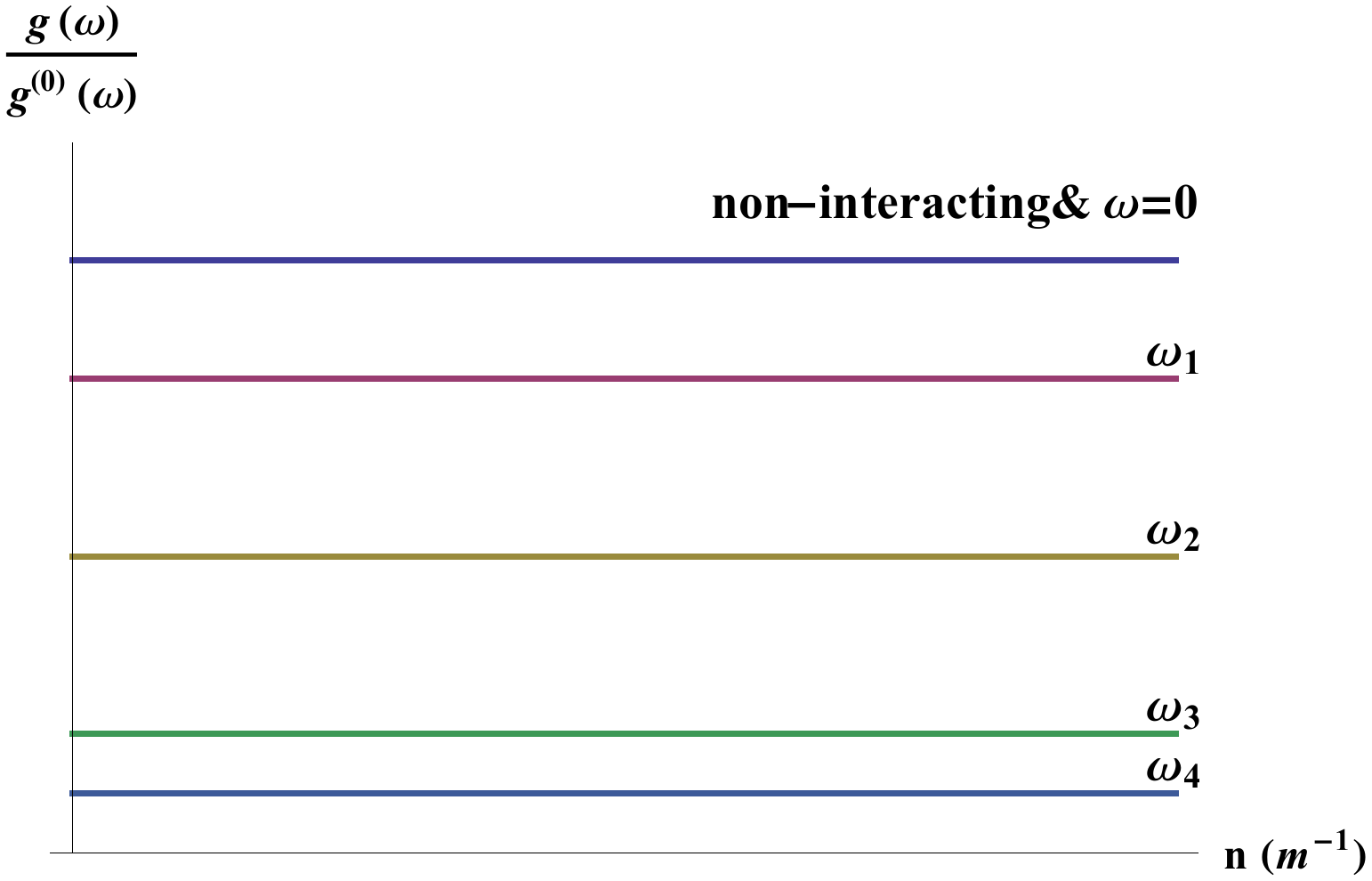}
\caption{(color online) \small{Plot of density of states normalized to the non-interacting value as a function of electron density \textit{n} for various values of $\omega$ ($\omega_1<\omega_2<\omega_3<\omega_4$). As we get further away from the Fermi points, the value of $g(\omega)$ decreases but remains constant for all density.}}
\end{figure}
\begin{figure}[h]
\centering
\includegraphics[scale=.5]{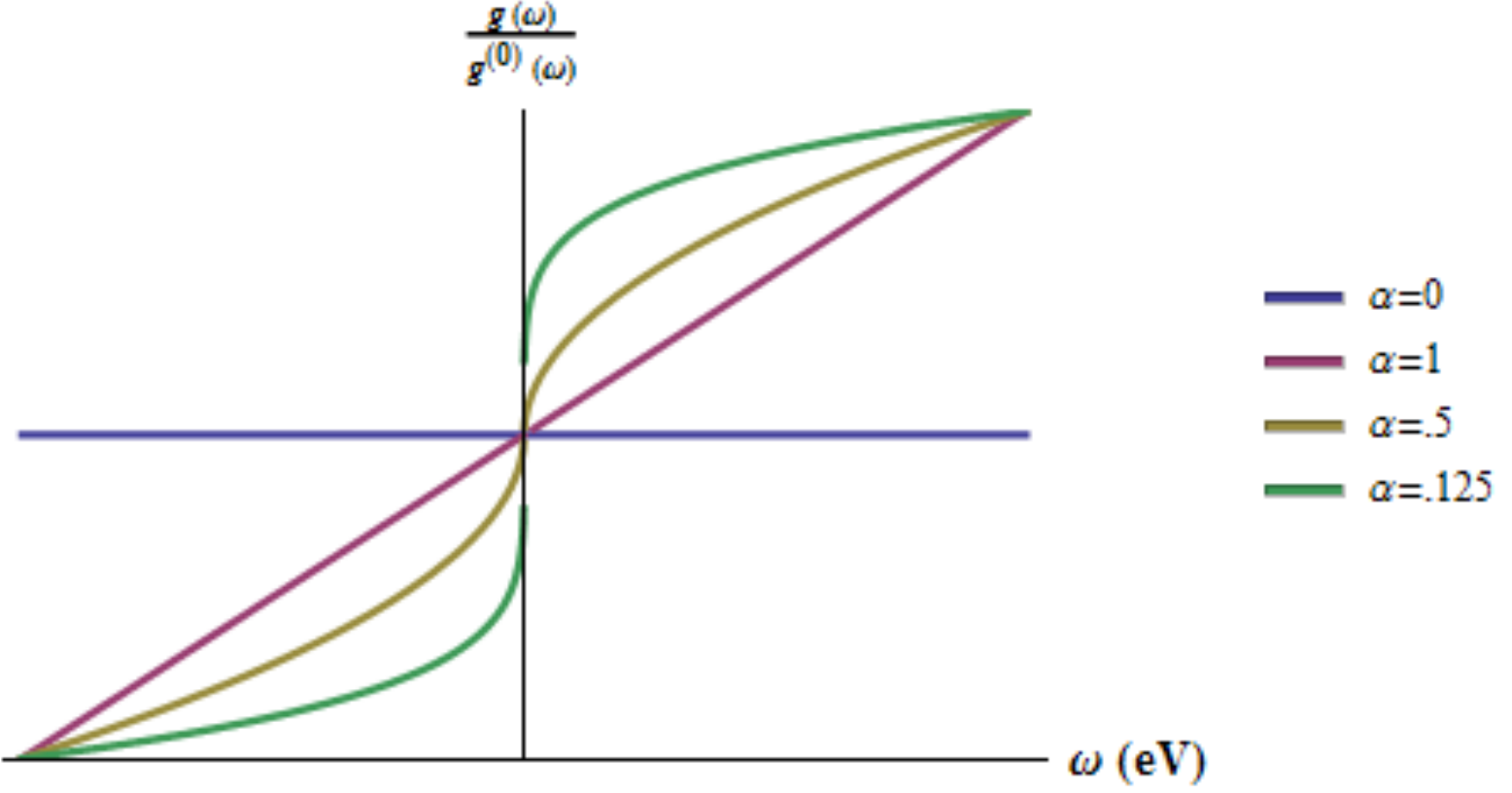}
\caption{(color online)\small{ Plot of density of states normalized to the non-interacting value as a function of frequency for various values of $\alpha$ (\cite{Ishii} obtained $\alpha=0.46\pm0.10$). In the presence of interactions ($\alpha\neq 0$), $g(\omega)$ behaves very differently from its non-interacting expression. We see from the above figure the predicted suppression of the density of states is seen in both normal metals and Dirac systems; $g/g^{(0)}\rightarrow0$ as $\omega$ increases.}}
\end{figure}
The density independence can also be seen in the collective modes and equilibrium properties of the system. Although these properties could be determined using eqns (13) and (14), we use the benefits of the reduced dimensionality and bosonize \cite{Mahan,Phillips,Giamarchi,Gogolin} the 1D Dirac material. The bosonized Hamiltonian has the following form:
\begin{equation}
H=H_\rho+H_\sigma
\end{equation}
which shows the separation of charge and spin that's seen in 1D interacting Fermi systems. These two independent boson modes give rise to the following dispersions:
\begin{equation}
\varepsilon_i=u_ik\hspace{.2in}i=\rho,\sigma
\end{equation}
where $u_\rho$ and $u_\sigma$ are the same as they were in eqn (10). A key feature in eqn (16) is the agreement with the energy derived from the virial theorem (eqn (4)). We readily see this given the following:
$$
u_{\rho,\sigma}^2=\left[v_g+\frac{1}{2\pi}\left(g_4^s\pm g_4^a\right)\right]^2-\left[\frac{1}{2\pi}\left(g_2^s\pm g_2^a\right)\right]^2
$$
and in the weak coupling limit:
$$
u_{\rho}^2=v_g^2+\frac{2}{\pi}v_gg_4\hspace{.25in}\&\hspace{.25in}u_{\sigma}=v_g
$$
These velocities, $u_\rho$ and $u_\sigma$ are independent of density. The only density dependence in the dispersion given by eqn (16) is in $k_F\propto\left(ar_s\right)^{-1}$ which agrees with our virial theorem result eqn (4). In the absence of a magnetic perturbation and the presence of a charge perturbation, the charge density mode propagates and gives rise to the speed of sound:
\begin{equation}
v_1=v_g\sqrt{1+\frac{2g_4}{v_g\pi}+\frac{1}{\left(v_g\pi\right)^2}\left[(g_4)^2-(g_2)^2\right]}
\end{equation} 
where we set $g_i=g_i^s=g_i^a$ based on anti-symmetry arguments. The speed of sound can also be obtained from the chemical potential as shown in \cite{Pines}. Both expressions are in agreement which shows we haven't missed any of the Hilbert space working in the boson representation \cite{Giamarchi}. In the weak coupling limit, we can neglect terms quadratic and higher in the coupling constant to obtain:
\begin{equation}
v_1=v_g\sqrt{1+\frac{2g_4}{v_g\pi}}
\end{equation}
which is the same expression obtained in \cite{Dzyalo}. Both expressions for the speed of sound, (17) and (18), have the predicted density independent behavior. Density independent speed of sound is surprising due to its counter-intuitive nature. Speed of sound (first sound in the Landau and Lifshitz language \cite{Landau}) is understood as a wave of compression and rarefaction. With an increase in electron density, one would expect a change in the speed of sound yet our result shows otherwise.

With the dispersion in (16), we can calculate all thermodynamic quantities. For example, the specific heat:
\begin{equation}
c_v(T)=\left[\frac{\pi k_B^2}{3}\left(\frac{1}{u_\sigma}+\frac{1}{u_\rho}\right)\right]T
\end{equation} 
has a linear dependence in temperature,  which is characteristic of both fermi and bose systems at low temperature. However, the difference appears in the bracketed term multiplying $T$ in (15). For typical 1D fermi systems, the term in brackets will depend inversely with density; in a SWNT the constant has a fixed value regardless of density. Thus the specific heat is also density independent. Other quantities such as the spin susceptibility, compressibility,     and conductivity can be derived using the bosonic dispersions \cite{Solyom,Voit,Giamarchi} and all present this density independence that's unique to 1D Dirac materials. In general, we conclude that due to the density independence in (8) and (14), all equilibrium quantities in 1D Dirac materials will themselves be independent of density in spite of the contradiction to conventional interpretation.

\section{Experimental Realization}
\label{Exp}
In order to test the density independent behavior predicted in this letter, we need scanning tunneling microscopy (STM). With STM, data for the density of states can be obtained  through \cite{Kane}:
\begin{equation}
\text{DoS}\propto\frac{dI}{dV}\approx V^\alpha
\end{equation}
where $\alpha$ has the same meaning as before,  \textit{V} is the voltage, and \textit{I} is the current.  Although many STM experiments have been done before on SWNTs, none had the goal of testing the density independence; previous experiments were focused on imaging and the dependence of the chiral indices on the density of states. In a sense, testing for this density dependence is easier than previous experiments due to the lifting of the uniformity requirement; we want multiple SWNTs with varying density. With that in mind, we expect the $dI/dV$ plots to remain independent of density. Prior experiments \cite{Ishii,Bockrath} have observed the temperature dependence of the differential conductance and have obtained good agreement with the theory. We would like to see results for keeping the temperature constant, and low, while changing the eletron density. We expect results similar to what can be seen in Fig. 3. In addition to revealing the density dependence in SWNTs, an experiment like this has the added bonus of re-affirming the existence of Luttinger-Liquid like states. Past experiments have been in agreement with theory, yet skeptics believe the observed behavior to be circumstantial due to the confined dimension of the electrons \cite{Deshpande}. Data from the proposed experiment testing density dependence would help solidify the idea that these states exist. Additionally, thermodynamic quantities (e.g. the specific heat) of SWNTs can be obtained at varying densities, with all other things constant, and should be the same value indicative of this electron density independent behavior.

\section{Concluding Remarks}
\label{Con}
In this work we have demonstrated that the exact linear dispersion and the constant Fermi velocity found in Dirac materials gives rise to nontrivial physics. By first using the virial theorem, we derive a closed form for the ground state energy with interactions included. Such an expression is rare and opens the door to calculation of the equilibrium quantities in the 1D Dirac material. Additionally, a strange similarity between TLL systems and Coulomb systems emerges that hints at the linearity of the dispersion, rather than the specific potential, as being more important in the unusual observed properties. We also observe some familiar 1D behavior, for example in the continuous momentum distribution function and low energy density of states, there are also profound differences. Primarily, the density independent exponents, equilibrium quantities, and collective modes, are differences that are unique to Dirac materials. These stark differences hint at a new class of 1D systems separate from traditional Tomonaga-Luttinger liquids. From studying these Dirac liquids we have a deeper understanding of the physics contained within Dirac materials; specifically,  the importance of interactions in the system, and the importance of dimensionality. From here, we hope to use this work to shed light on even more exotic Dirac materials. 

\begin{acknowledgements}
The authors, M.P. Gochan and K.S. Bedell, would like to thank Thomas Mion and Dr. Ilija Zeljkovic for valuable discussion in regards to experimental techniques as well as James Stokes, and Dr. Alexander Balatsky, for fruitful discussion. Additionally, we thank the IMS at LANL for their hospitality. This work was supported in part by the John H. Rourke, Boston College, Endowment Fund and the NSF Grant PHY-1208521
\end{acknowledgements}


\bibliographystyle{spmpsci}
\bibliography{One_Dim_DM}
\end{document}